# Mechanically-Controlled Binary Conductance Switching of a Single-Molecule Junction

Su Ying Quek[1], Maria Kamenetska[2,3], Michael L. Steigerwald[4], Hyoung Joon Choi[5], Steven G. Louie[1,6], Mark S. Hybertsen[7], J. B. Neaton[1], L. Venkataraman[2,3*]

[1]Molecular Foundry, Lawrence Berkeley National Laboratory, Berkeley, CA
[2]Department of Applied Physics and Applied Mathematics, Columbia University, New York, NY
[3]Center for Electron Transport in Nanostructures, Columbia University, New York, NY
[4]Department of Chemistry, Columbia University, New York, NY
[5]Department of Physics and IPAP, Yonsei University, Seoul, Korea
[6]Department of Physics, University of California, Berkeley, Berkeley, CA
[7]Center for Functional Nanomaterials, Brookhaven National Laboratory, Upton, NY

AUTHOR EMAIL ADDRESS: JBNeaton@lbl.gov; lv2117@columbia.edu

Molecular-scale components are expected to be central to nanoscale electronic devices[1-3]. While molecular-scale switching has been reported in atomic quantum point contacts[4-6], single-molecule junctions provide the additional flexibility of tuning the on/off conductance states through molecular design. Thus far, switching in single-molecule junctions has been attributed to changes in the conformation or charge state of the molecule[7-12]. Here, we demonstrate reversible binary switching in a single-molecule junction by mechanical control of the metal-molecule contact geometry. We show that 4,4'-bipyridine-gold single-molecule **junctions can be reversibly switched between two conductance states through repeated junction elongation and compression. Using first-principles calculations, we attribute the different measured conductance states to distinct contact geometries at the flexible but stable N-Au bond: conductance is low when the N-Au bond is perpendicular to the conducting π-system, and high otherwise. This switching mechanism, inherent to the pyridine-gold link, could form the basis of a new class of mechanically-activated single-molecule switches.**

In this work, we focus on 4,4'-bipyridine-gold junctions, where we find two reproducible and distinct conductance states that can be controllably switched by mechanical manipulation of the electrode separation. The conductance is measured by repeatedly forming and breaking Au point contacts with a modified STM in a solution of the molecules at room temperature[13, 14]. The current is recorded at a fixed bias while the junction is elongated to generate conductance traces (see Methods). Conductance histograms are constructed from thousands of traces where peaks correspond to the most frequently observed conductance values. With this method, single molecule junction conductances can be measured reliably and reproducibly for molecules with amine[14, 15], methyl sulfide, and phosphine[16] link groups.

Fig. 1a shows a normalized conductance histogram for bipyridine-Au junctions determined from over 10,000 consecutively measured conductance traces without any data selection or processing, along with a histogram measured in solvent alone. Our histogram shows two clear peaks, corresponding to the most frequently observed conductance values, centered around $1.6 \times 10^{-4}$ $G_0$ (Low G) and $6 \times 10^{-4}$ $G_0$ (High G), and a high conductance tail is seen extending to ~$3 \times 10^{-3}$ $G_0$, where $G_0 = 2e^2/h$, is the quantum of conductance. These differ from previous reported measurements for bipyridine-Au junctions[13, 17] where histograms were constructed from a smaller number of selected traces and the measured conductances, encompassing the High G and High G tail region, did not overlap with the entire range studied here.

Typical conductance traces, shown in Fig. 1B, exhibit a High G step that precedes a Low G step. To statistically analyze this step sequence in our entire data set, we compute a two dimensional (2D) conductance-displacement histogram of all measured traces (see Methods). The 2D histogram generated from the same 10000 traces (Fig. 1c) shows two clear regions with a large number of counts. The counts within the Low G range occur ~2 Å after the break of the gold point contact (x=0). This is in contrast to the counts in the High G range, which start right after the break of the gold point-contact. This indicates that the High G steps start as soon as the gold contact breaks, and Low G steps follow High G steps (see SI for more details).

Since the Low G steps occur only upon elongation of the junction, a natural question is whether junction compression would restore the High G state. To investigate this possibility, we measure the conductance between the tip and substrate while applying two types of ramps (dashed trace in Fig. 2a and SI Fig. 2a) to the piezoelectric actuator that modulates the substrate position (while the tip is held fixed) (See Methods). When a molecule is present, switching events between the high and low conducting states are frequently seen. Sample switching conductance traces in Figs. 2a and SI Fig. 2a clearly show the junction is controllably modulated between the high and low conducting states as the tip-sample separation is modulated. Out of 9000 total traces with each ramp, we find that a molecule is present in the Low G state about ~10% of the time after the initial 15 Å displacement. Among these 1057 traces, ~90% had at least one switching event, and ~ 20% had switching events that lasted through the entire sequence, as shown. Fig. 2b (and SI Fig. 2b) show companion conductance histograms constructed from these selected traces. Like Fig. 1a, these histograms show two clear peaks indicating that the molecular junction predominantly samples High G and Low G conductances when successively compressed and elongated. These measurements thus illustrate that a bipyridine-Au junction can be switched between two well-defined conductance states by mechanical control of the tip-sample distance. Control experiments on the solvent alone, 3,3',5,5'-Tetramethyl 4,4'Diamino-Biphenyl, 2,4'-bipyridine and 4-phenyl-pyridine (see SI) show that this switching behavior is intrinsic to pyridine-Au bonds present at both ends of the molecule. Furthermore, we find that a 2-3Å piezo modulation is required to switch the molecule (SI Fig. 5).

From Fig. 1c we know that bipyridine junctions form right after breaking a Au-Au point contact, where the tip-sample distance has been experimentally shown[18] to be around 6.5 Å (± 2.5Å), shorter than the N to N distance along bipyridine of 7.2 Å (See SI Fig. 6). Further, allowing for the length of two N-Au bonds (~2.1 Å each), the as-formed electrode structures will frequently impose strong geometric constraints on the initial junction geometries. Using a modified piezo ramp as shown in Fig. 2c, we correlate junction conductance to the distance required to push back the electrodes together to form a Au-Au contact (see Methods). This "push-back" distance, a measure of the electrode separation for the junction, is shown in Fig. 2d as a function of conductance, for 777 traces. Junctions with a conductance in the Low G range have a push-back distance of greater than 10 Å, consistent with a geometry where the molecule is held vertically between the two electrodes, and corresponding to an Au-Au electrode separation of around 12-13 Å (estimating the Au-Au separation at contact using the (111) interplanar spacing). Junctions with a conductance in the High G range have a push-back distance which decreases from 10 Å to about 5 Å as the conductance increases from ~$3\times10^{-4}G_0$ to $4\times10^{-3}G_0$.

The properties of the pyridine-gold link naturally explain the observed switching behavior. The bonding mechanism, elucidated by our density functional theory (DFT) calculations detailed below, consists of donation from the N lone pair orbital into the partially empty s-orbital on a specific undercoordinated Au atom on the electrode. Since the N lone pair in bipyridine is parallel to the bipyridine backbone, we expect the N-Au bond to be along the bipyridine backbone. While such a structure is difficult to achieve initially given the geometric constraints, it may be easily accommodated after elongation by several Å. Previous conductance calculations[19-21], in agreement with our own, have shown that the essential orbital channel supporting transmission is the lowest unoccupied π*-orbital (LUMO; Fig. 3a). Since the π*-orbital is orthogonal to the N lone pair in this case, it is plausible to expect that an elongated junction, with the N-Au bond aligned to the backbone, will have low electronic coupling and hence low conductance. On the other hand, the constraints imposed by the compressed junctions will drive strong tilting of the N-Au bond, which can result in stronger coupling and higher conductance.

Our DFT calculations (see Methods) indicate that bipyridine molecules bind selectively to undercoordinated atop Au sites. To investigate the sensitivity of conductance to N-Au bond orientation, we compute the transmission for a series of model junctions (Fig. 3b) with identical geometric features except for the angle $\alpha$ between the N-Au bond and $\pi^*$-system (Fig. 3a). Using a prototypical relaxed junction with a vertical molecule ($\alpha = 90°$) (Fig. 3b, panel 1), we obtain a self-energy corrected[22] (see Methods, SI) transmission (Fig. 3c). As mentioned above, the LUMO $\pi^*$-state (Fig. 3a) provides the dominant contribution to the conductance.

The width of the LUMO-derived transmission peak increases almost linearly with $\cos(\alpha)$ (Fig. 3c), reflecting an enhanced electronic coupling between the Au $s$-state and the LUMO $\pi^*$-orbital as the N-Au bond tilts out of the plane of the pyridine ring. This enhanced coupling also leads to an increased back-donation of electrons to the molecule, raising its local electrostatic potential and shifting the molecular levels to higher energies. Since the transmission at $E_F$ is related to the tail of a resonance, the conductance trend is dominated by its width, and the low bias conductance increases with N-Au bond tilt (Fig. 3c inset). Tilting the N-Au bond out of the pyridine plane does result in a decreased junction binding energy, from 1.36 eV for a vertical junction ($\alpha = 90°$), to 1.03 eV for $\alpha = 50°$, and to 0.70 eV for $\alpha = 30°$. Thus, at modest cost in binding energy, the local metal-molecule contact geometry can vary substantially with corresponding large changes in junction conductance.

The versatile amine-Au link chemistry is also based on selective donor-acceptor bonding. In the amine case, the amine lone pair is naturally coupled into the main orbital responsible for conductance e.g. into the benzene $\pi$-system or the alkane $\sigma$-system[23]. Changes in contact geometry do not, therefore, affect the measured and calculated conductance significantly[22], resulting in a single, well defined peak in the conductance histograms[14], and no mechanically-induced switching behavior (SI Fig. 4). In contrast, for bipyridine-Au junctions, the N lone pair electrons that dominate bond formation are actually *orthogonal* to the π-system resulting in bipyridine junction conductances that can be quite sensitive to the orientation of the N-Au bond relative to its principal conducting orbital, the molecular LUMO.

We consider a total of 55 relaxed junctions (see SI), small compared with the experimental sample size but sufficient to explore the impact of junction geometry on conductance. To model tractably the local roughness and large radius of curvature expected for initially broken soft Au contacts, we consider relaxed junctions in which bipyridine is bonded to one- and two-layer Au motifs (adatom, dimer, trimer, pentamer, pyramid) on Au(111). Junctions with tip-sample distances close to that of the initially-broken Au contact have more constrained geometries (smaller $\alpha$) and higher conductances (Fig. 4a structures 1, 2), while larger electrode-electrode separations accommodate geometries with a larger $\alpha$ and lower conductances (Fig. 4a structures 3, 4). The calculated conductance (with self-energy corrections) is plotted as a function of the vertical distance between Au binding sites in Fig. 4b. Comparing this to the corresponding experimental plots (Figs. 1c and 2d) reveals good quantitative agreement between the predicted and measured conductance ranges as well as their relation to Au-Au separation.

In Fig. 4c, we plot the calculated conductance as a function of $\alpha$ which shows a conductance increase with decreasing $\alpha$, as discussed above. We see that conductance is also affected by the N-Au bond lengths, the degree of coordination at the binding sites,[19] and the torsional angle between the rings (which varies from ~21-42° in the relaxed geometries). For junctions with conductance in the high G tail of the experimental histogram (> ~1×10$^{-3}$ $G_0$ in Fig. 1a), the LUMO was observed to have significant overlap with orbitals on adjacent Au contact atoms. This extra coupling is due an additional broadening of the LUMO (not a new conducting channel) and is controlled by the separation between the nearby Au electrode atom and one of the C atoms in the pyridine ring. In Fig. 4d, the conductance is plotted as a function of the minimum C-Au distance (d(C-Au)) between a C atom in the molecule and Au atoms on the electrode. The cluster of points with minimum d(C-Au) < ~2.8 Å have conductances > $1 \times 10^{-3}$ $G_0$. In our calculations, we find that geometries with conductance in the experimental High G range have small $\alpha$ (< ~70°) and/or small minimum d(C-Au).

These results guide us in proposing a working hypothesis for the distinct High G and Low G steps observed in our experiments. An initially-broken Au contact has a Au-Au separation that is too small to accommodate a bipyridine molecule in a vertical geometry (Fig. 4e). Junctions formed at the beginning of a pull trace thus have geometries that result in a high conductance. As the junction is elongated, the High G geometry tends to snap to a Low G geometry, once binding sites spaced far enough apart become available (Fig. 4f). This is plausible given that the Au contacts in the experiment are likely to offer multiple binding sites, as illustrated in Figs. 4e and 4f. On the other hand, as the junction is compressed from a vertical geometry, the energy cost associated with rotating the N-Au bond out of the plane of the pyridine backbone tends to hold $\alpha$ ~90°. Only when the constraints of the shrinking junction demand too much bond compression will the Low G geometry snap to a high G geometry with smaller $\alpha$ or C-Au distance. In this picture, the N-Au bonds need not be broken, although an Au contact atom may shift from one electrode site to another. It is also plausible that the bond may shift from one available undercoordinated Au site to another, as illustrated in Figs. 4e and 4f. In either case, although not strictly reversible in all the atomic scale details, this picture provides the essential elements for a mechanically activated switch.

**Methods**:

Experimental Methods: We measured the molecular conductance of 4,4' Bipyridine (Sigma-Aldrich, 98% purity) by repeatedly forming and breaking Au point contacts in solution of the molecules with a home-built, simplified STM (see Supplementary document for details). Thousands of traces are collected and presented as conductance histograms, where peaks correspond to the most frequently observed conductance values. A freshly cut gold wire (0.25 mm diameter, 99.999% purity, Alfa Aesar) was used as the tip, and UV/ozone cleaned Au substrate (mica with 100 nm Au, 99.999% purity, Alfa Aesar) was used as the substrate. The STM operates in ambient conditions at room temperature and the junctions were broken in a dilute, 1mM, solution of 4,4' Bipyridine in 1,2,4-trichlorobenzene (Sigma-Aldrich, 99% purity). To ensure that each measurement started from a different initial atomic configuration of the electrodes, the electrodes were pulled apart only after being brought into contact with the Au surface, indicated by a conductance greater than a few $G_0$. Prior to adding a molecular solution between the tip and substrate, 1000 conductance traces were first collected without molecules to ensure that there were no contaminations in the STM set-up.

For all non-linear ramps applied to the piezo, the junction was first closed to achieve a conductance larger than a few $G_0$. For demonstrating switching between the high and low conducting states (Figure 2a), the junction was then pulled apart by 15 Å at 16 nm/s, and then sequentially pushed together and pulled apart by 2 Å four times, holding the junction for 15 ms at each step, before the junction was finally extended by an additional 25 Å and broken. For the measurement of the "push-back" distance, the junction was first pulled apart by 15 Å at 16 nm/s, held at this separation for 50 ms, pushed together by 15 Å at 16 nm/s, and then pulled apart and broken. This cycle was repeated 2000 times and the measured conductance data was analyzed to determine the "push-back" distance as follows. For each measured trace, we determine the average conductance while the electrodes were held fixed (second segment of the ramp), and the distance the junction had to be pushed together before reaching a conductance within the range of that expected for a Au-Au contact (as illustrated with red and blue arrows in Fig 2c). Of the 2000 traces measured, 777 had a conductance between $1\times10^{-4}$ $G_0$ and $1\times10^{-2}$ $G_0$. We divided the entire conductance range into 20 bins in the logarithmic scale and averaged the conductance and push-back distance for all points within each conductance bin to obtain the data shown in Figure 2d.

Construction of two-dimensional histogram: Each measured conductance trace consists of conductance data acquired every 25 μs, measured as a function of tip-sample displacement at a constant 16 nm/s velocity. Since gold and molecular conductance plateaus occur in random locations along the entire displacement axis (x-axis) within the measured range, we first set the origin of our displacement axis at the point in the conductance traces where the gold-gold contact breaks and the conductance drops below $G_0$. This well-defined position on the x-axis is determined individually for each trace as illustrated in the inset of Fig. 1c using an unbiased automated algorithm[22]. For about 1% of the measured traces, this position cannot be determined and these traces are not used for further analysis. Each data point on the digitized conductance trace now has a conductance coordinate (along the y-axis) and a position coordinate (along the x-axis).

These data are binned using a linear scale along the displacement axis and a log-scale along the conductance to generate a 2D histogram.

Transport Calculations: First-principles transport calculations are based on density functional theory (DFT) within the generalized gradient approximation (GGA).[24] The SCARLET code[25] is used to calculate the electron transmission for many junction geometries. The linear response conductance is obtained from the Landauer formula (G = $T(E_F)*G_0$), where T(E) is the transmission function, and $E_F$ is the Fermi energy. The alignment of the frontier molecular energy levels in the junction relative to $E_F$ can show significant errors in DFT[26] with the result that the calculated conductance are too large[22, 27-29]. A self-energy correction, successfully used in our previous work[22], is calculated and added to the molecular orbital energies in the junction to account for many-electron effects. Details of the application to the bipyridine case are described in the Supplementary materials. Because the DFT orbital energy is close to the electrode Fermi energy in this system, the self-energy correction is quantitatively quite important to the predicted conductance value.

**Acknowledgments:**
We thank Chris Wiggins and Philip Kim for discussions. Portions of this work were performed at the Molecular Foundry, Lawrence Berkeley National Laboratory, and were supported by the Office of Science, Office of Basic Energy Sciences, of the U.S. Department of Energy. This work was supported in part by the Nanoscale Science and Engineering Initiative of the NSF (award numbers CHE-0117752 and CHE-0641532), the New York State Office of Science, Technology, and Academic Research (NYSTAR) and the NSF Career Award (CHE-07-44185) (MK and LV). This work was supported in part by the US Department of Energy, Office of Basic Energy Sciences, under contract number DE-AC02-98CH10886 (MSH). H.J.C. acknowledges support from KISTI Supercomputing Center (KSC-2007-S00-1011). Computational resources from NERSC are acknowledged.

**Supplementary Information:**
Supplementary information accompanies this paper at
www.nature.com/naturenanotechnology. Reprints and permission information is
available online at http://npg.nature.com/reprintsandpermissions/.

**Fig. Captions:**

**Fig. 1: Statistical analysis of measured conductance traces. (a) Normalized conductance histogram for 4.4' bipyridine.** The histogram is constructed without any data selection from 10000 traces measured at a 25 mV bias voltage using a conductance bin size of $10^{-6}G_0$ along with a histogram collected in solvent alone (yellow). Black dashed lines show Lorentzian fits to the two peaks. Arrows indicate the High G and Low G peaks. Inset: Same histograms shown on a log-log scale using a bin size of $10^{-5}G_0$ **(b) Sample conductance traces measured at a 25 mV bias and 16 nm/s displacement speed showing two conductance steps in succession. (c) Inset: Sample conductance trace which demonstrates how the displacement origin was selected for each trace to construct the two-dimensional (2D) histogram in (c). (c) 2D histogram constructed from all traces with a clear $G_0$ break**. Two regions with a large number of counts, encircled by the black dashed lines, are clearly visible. The High G region, around $10^{-3}G_0$ extends from the origin to about 4Å along the x-axis and Low G region, around $3\times10^{-4}G_0$ start ~2Å displaced from the origin. Many High G steps exhibit some slope, as can be seen from the orientation of the High G region in the plot.

**Fig. 2**: **Controlled conductance switching by mechanical manipulation of Au-Au distances. (a) Sample bipyridine switching conductance traces (colored solid lines).** These traces were collected while applying the non-linear ramps (dashed black line) shown measured at a 250 mV applied bias. **(b) Conductance histograms of 1057 switching traces that had a molecule in the Low G state after the initial 15Å**

**displacement**. These histograms are constructed using conductance data from the ramp section of the trace only. Traces in (a) show reversible switching between conductance states that are around the two peaks clearly visible in the companion histogram. (Note: the peak positions are slightly shifted from those in Fig. 1 because of different experimental conditions and analysis method). **(c) Sample conductance traces (blue and red) measured while applying the non-linear ramp shown (grey trace).** The blue trace has a conductance in the Low G range during the "hold" section, while the red trace has a conductance in the High G range. Push-back distances are determined as shown by the blue and red arrows, using an automated procedure. **(d) Average conductance as a function of average push-back distance for 777 of 2000 traces measured (red ×).** (See Methods) Data shows that for junctions with a conductance in the Low G range, the push back distance is around 10-11 Å, while for junctions with a conductance in the High G range, the push-back distance increases with decreasing conductances. Error bars are one standard deviations in both conductances and push-back distance. Also shown is the conductance histogram from Fig. 1a (solid red line) along the same conductance axis.

**Fig. 3: Calculated transmission characteristics as a function of the angle between the N-Au bond and the π∗-system. (a) Schematic showing the coupling between the Au-s orbital (orange) with the bipyridine LUMO.** $\alpha$ denotes the angle between the N-Au bond and the π∗-system. **(b) Junction geometries of bipyridine bonded on each side to Au adatoms on Au(111), with varying $\alpha$ (labeled in Fig.). (c) Self-energy corrected transmission functions plotted on a semi-log scale for junctions in (b).** Black solid, red dashed, blue dashed-dotted, green dotted lines denote $\alpha$ = 90°, 70°, 50° and 30° respectively. The inset shows G, given by $T(E_F)*G_0$, decreasing with increasing $\alpha$.

**Fig. 4: Results from conductance calculations on 55 relaxed junctions. (a) Examples of junction geometries relaxed at different tip-sample distances. (b-d) Self-energy corrected conductance G for 55 relaxed junctions, plotted against (b) the vertical distance between Au contacts, (c) the angle $\alpha$ between the N-Au bond and π-system (as illustrated in Figure 3), and (d) the minimum C-Au distance.** The series of points for $\alpha = 90°$ corresponds to different N-Au bond lengths and binding sites in a vertical junction. Despite the spread, they all fall within the experimental Low G range. **(e, f) Schematic illustrating the High G and Low G configurations respectively that exhibit mechanically-induced switching for junctions highlighting the role of the geometric constraints and Au tip morphology.**

**References**:

1. Joachim, C., Gimzewski, J. K. & Aviram, A. Electronics using hybrid-molecular and mono-molecular devices. Nature 408, 541-548 (2000).
2. Mathur, N. Nanotechnology - Beyond the silicon roadmap. Nature 419, 573-575 (2002).
3. Nitzan, A. & Ratner, M. A. Electron transport in molecular wire junctions. Science 300, 1384-1389 (2003).
4. Smith, D. P. E. Quantum Point-Contact Switches. Science 269, 371-373 (1995).
5. Terabe, K., Hasegawa, T., Nakayama, T. & Aono, M. Quantized conductance atomic switch. Nature 433, 47-50 (2005).
6. Xie, F. Q., Nittler, L., Obermair, C. & Schimmel, T. Gate-Controlled Atomic Quantum Switch. Physical Review Letters 93, 128303 (2004).
7. Moresco, F. et al. Conformational changes of single molecules induced by scanning tunneling microscopy manipulation: A route to molecular switching. Physical Review Letters 86, 672-675 (2001).
8. Chen, F. et al. A molecular switch based on potential-induced changes of oxidation state. Nano Letters 5, 503-506 (2005).
9. Blum, A. S. et al. Molecularly inherent voltage-controlled conductance switching. Nature Materials 4, 167-172 (2005).
10. Lortscher, E., Ciszek, J. W., Tour, J. & Riel, H. Reversible and controllable switching of a single-molecule junction. Small 2, 973-977 (2006).

11. Li, X. L. et al. Controlling charge transport in single molecules using electrochemical gate. Faraday Discussions 131, 111-120 (2006).
12. Liljeroth, P., Repp, J. & Meyer, G. Current-induced hydrogen tautomerization and conductance switching of naphthalocyanine molecules. Science 317, 1203-1206 (2007).
13. Xu, B. Q. & Tao, N. J. J. Measurement of single-molecule resistance by repeated formation of molecular junctions. Science 301, 1221-1223 (2003).
14. Venkataraman, L. et al. Single-Molecule Circuits with Well-Defined Molecular Conductance. Nano Letters 6, 458 - 462 (2006).
15. Venkataraman, L., Klare, J. E., Nuckolls, C., Hybertsen, M. S. & Steigerwald, M. L. Dependence of single-molecule junction conductance on molecular conformation. Nature 442, 904-907 (2006).
16. Park, Y. S. et al. Contact Chemistry and Single Molecule Conductance: A Comparison of Phosphines, Methyl Sulfides and Amines. J. Am. Chem. Soc. 129, 15768-15769 (2007).
17. Zhou, X. S. et al. Single molecule conductance of dipyridines with conjugated ethene and nonconjugated ethane bridging group. Journal of Physical Chemistry C 112, 3935-3940 (2008).
18. Yanson, A. I., Bollinger, G. R., van den Brom, H. E., Agrait, N. & van Ruitenbeek, J. M. Formation and manipulation of a metallic wire of single gold atoms. Nature 395, 783-785 (1998).
19. Stadler, R., Thygesen, K. S. & Jacobsen, K. W. Forces and conductances in a single-molecule bipyridine junction. Physical Review B 72, 241401 (2005).
20. Perez-Jimenez, A. J. Uncovering transport properties of 4,4 '-bipyridine/gold molecular nanobridges. Journal of Physical Chemistry B 109, 10052-10060 (2005).
21. Hu, Y. B., Zhu, Y., Gao, H. J. & Guo, H. Conductance of an ensemble of molecular wires: A statistical analysis. Physical Review Letters 95, 156803 (2005).
22. Quek, S. Y. et al. Amine-Gold Linked Single-Molecule Junctions: Experiment and Theory. Nano Letters 7, 3477-3482 (2007).
23. Hybertsen, M. S. et al. Amine-linked single-molecule circuits: systematic trends across molecular families. Journal of Physics: Condensed Matter 20, 374115 (2008).
24. Perdew, J. P., Burke, K. & Ernzerhof, M. Generalized gradient approximation made simple. Physical Review Letters 77, 3865-3868 (1996).
25. Choi, H. J., Marvin, L. C. & Steven, G. L. First-principles scattering-state approach for nonlinear electrical transport in nanostructures. Physical Review B (Condensed Matter and Materials Physics) 76, 155420 (2007).
26. Neaton, J. B., Hybertsen, M. S. & Louie, S. G. Renormalization of molecular electronic levels at metal-molecule interfaces. Physical Review Letters 97, 216405 (2006).
27. Koentopp, M., Burke, K. & Evers, F. Zero-bias molecular electronics: Exchange-correlation corrections to Landauer's formula. Physical Review B 73, 121403 (2006).

28. Toher, C. & Sanvito, S. Efficient atomic self-interaction correction scheme for nonequilibrium quantum transport. Physical Review Letters 99, 4 (2007).
29. Ke, S. H., Baranger, H. U. & Yang, W. T. Role of the exchange-correlation potential in ab initio electron transport calculations. Journal of Chemical Physics 126, 4 (2007).

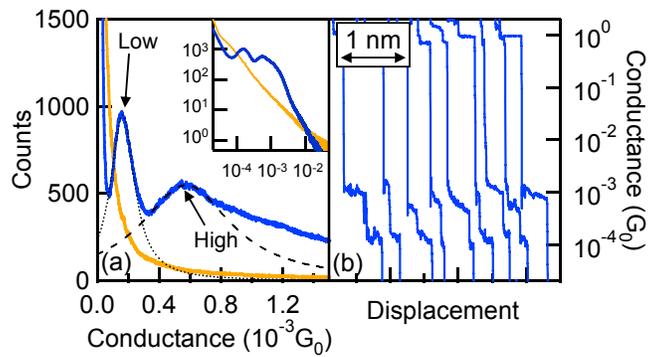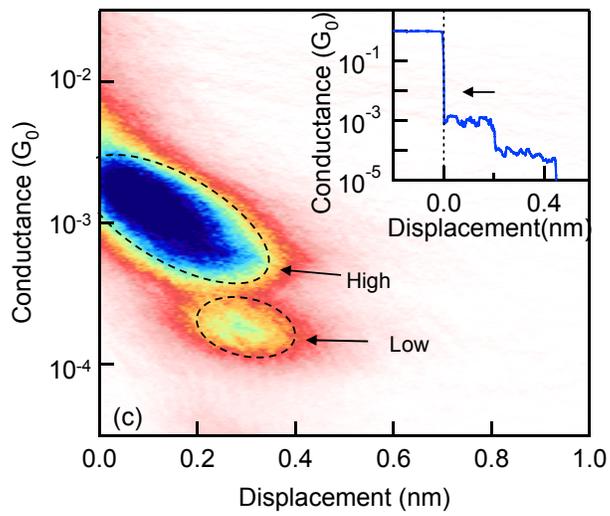

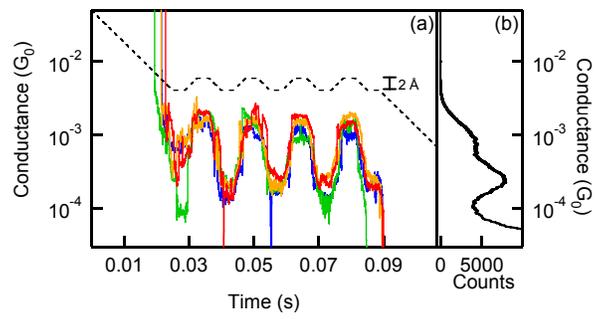
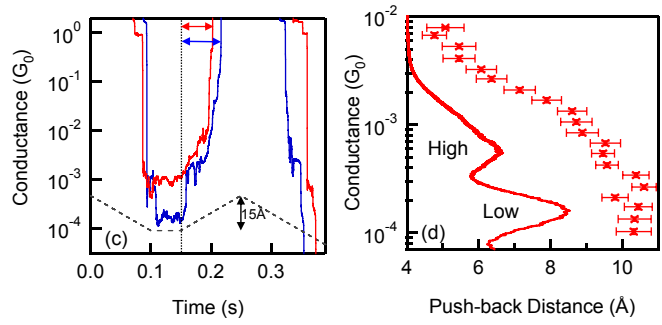

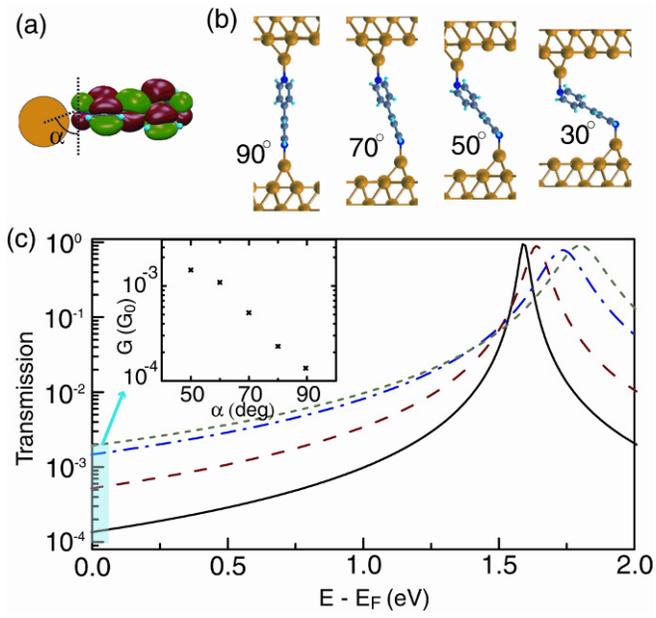

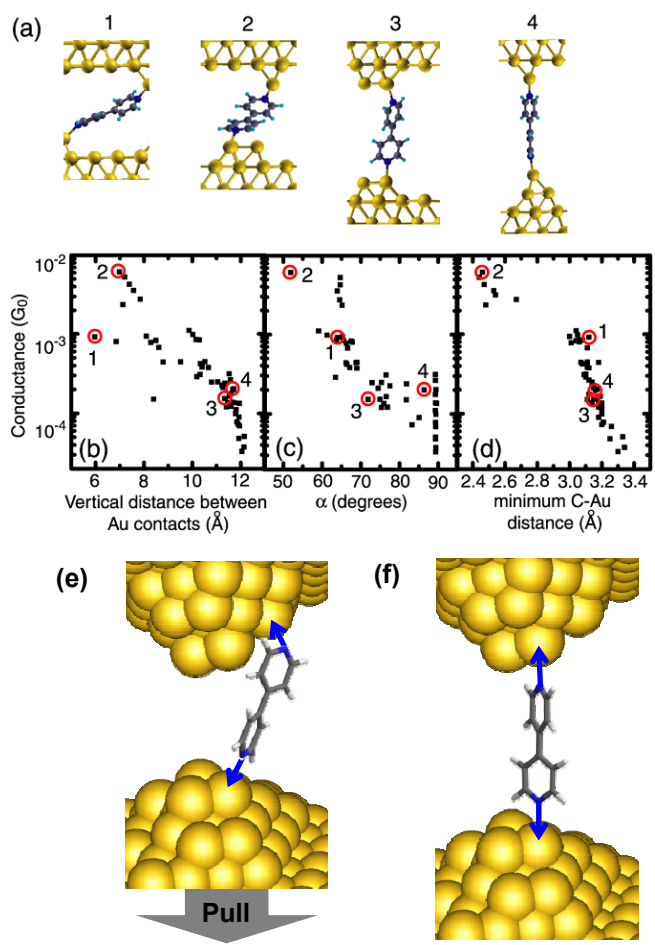